# Dynamic vortical flow profile quantitatively characterizes mitral valvular-left ventricular hemodynamic coupling: In vivo analysis by dynamic 3D enstrophy mapping from 4D Flow MRI

Mohammed S.M. Elbaz, PhD[1,2*], Trung Bao Le, PhD[3,7], Pankaj Garg, MD, PhD[4],
Arno A.W. Roest, MD, PhD[5], Boudewijn P.F. Lelieveldt, PhD[2,6], Sven Plein, MD, PhD[4],
Jos J.M. Westenberg, PhD[2], Fotis Sotiropoulos, PhD[3], Rob J. van der Geest, PhD[2]

[1] Department of Radiology, Northwestern University, Chicago, Illinois, USA [2] Department of Radiology, Leiden University Medical Center, Leiden, The Netherlands [3] Department of Civil Engineering, Stony Brook University, Stony Brook, NY, USA [4] Leeds Institute of Cardiovascular and Metabolic Medicine (LICAMM), University of Leeds, Leeds, UK [5] Department of Pediatrics, division of Pediatric Cardiology, Leiden University Medical Center, Leiden, The Netherlands [6] Pattern Recognition and Bioinformatics Group, Faculty of Electrical Engineering, Mathematics, and Computer Science, Delft University of Technology, Delft, The Netherlands [7] Computational Fluids Laboratory, North Dakota State University, USA

**Abstract**

Vortical blood flow in the human left ventricular (LV) inflow initiates from the mitral valve (MV) and evolves within the LV during diastolic E-filling. Hence, vortical flow links MV and LV hemodynamics. This study sought to elucidate and quantitatively characterize the in vivo 3D dynamics of LV vortical flow over E-filling and relation to MV-LV hemodynamic coupling using 4D Flow MRI flow field. 34 healthy volunteers and 5 example patients underwent 4D Flow MRI. Vortical blood flow evolution was mapped in the LV over E-wave using enstrophy density. A new dimensionless profile $P_{MV-LV}$ was derived as a function of both MV vortex formation time (VFT) and LV volumetric enstrophy density. Results reveal that 3D vortical flow evolution in the healthy LV follows a bi-phasic behavior with a vortical growth phase followed by a vortical decay phase. In healthy LVs studied, the $P_{MV-LV}$ profile showed that the vortical growth and decay phases are characterized by a vortical growth time $T_{growth} = 1.23 \pm 0.25$, growth rate $\alpha = 0.80 \pm 0.17$, decay time $T_{decay} = 0.96 \pm 0.39$ and decay rate $\beta = -1.02 \pm 0.49$. Distinctly altered parameters were found in the pilot patients studied. The derived $P_{MV-LV}$ profile quantitatively characterizes MV-LV hemodynamic coupling by vortical flow dynamics. Results herein unravel new insights into cardiac physiology and could enable a novel standardized methodology to study MV-LV hemodynamic coupling and association to cardiac function in health and disease.

**Key terms**: vortex flow, cardiac function, heart physiology, cardiac hemodynamics, left ventricle, mitral valve, 4D Flow MRI, cardiac blood flow, hemodynamic coupling

*Correspondence to: mohammed.elbaz@northwestern.edu

**Introduction**

In-vivo and in-vitro studies have shown that the inflow of blood into the human heart's left ventricle (LV) during E-filling provokes a swirling motion forming distinct vortical flow patterns (1-6). LV vortical flow is thought to help to preserve momentum, minimize kinetic energy loss and to help redirecting mitral inflow blood towards the aortic outflow tract (2, 6, 7). The formation of vortical flow in the human heart has been extensively studied in-vivo and in-vitro (2-5, 7, 10-13). Altered vortical flow formation was reported in patients with congenital and acquired heart disease (1, 5, 8, 9) and was recently associated to elevated levels of intra-cardiac viscous energy loss (7, 10). Nevertheless, the role of vortical blood flow in cardiac (patho)physiology is not yet fully understood and remains a subject of debate in the literature (11-13).

Earlier studies have shown that diastolic vortical flow initiates, with the start of the E-filling, from the shear layers distal to the mitral valve (MV) leaflets as a result of MV inflow interaction with the MV geometry. Thus, vortical flow formation has been characterized as a function of MV geometry and inflow dynamics (1, 5, 14, 15). However, such characterization implies that LV vortical flow is uni-phasic dominated by vortical flow formation from the MV and does not take into account the impact of the dynamic LV geometry on vortical flow. Recent studies have shown that the formed vortical flow continues to evolve within the dynamic LV geometry over diastole resulting in time-varying vortical flow patterns (3, 4, 16, 17). Thus, suggesting an impact of the dynamic LV geometry on the evolution of vortical flow. While previous studies have provided a better understanding of the LV vortical flow formation phenomenon (1, 4-6, 16, 17), there is currently a lack of quantitative methods for characterizing or mapping of the time-evolution of vortical flow within the dynamic LV geometry of healthy subjects and patients. As a result, the actual in vivo evolution of vortical flow in the LV over E-filing remains unclear.

*Correspondence to: mohammed.elbaz@northwestern.edu

Given that vortical flow, during E-filling, emerges from MV inflow dynamics but progresses inside the LV bounded by its dynamics geometry (1, 3, 4), we hypothesized that vortical flow evolution would be determined as a function of both MV and LV dynamics (i.e. is a reflection of the coupled dynamics between MV and LV). Therefore, characterization of the 3D intra-cardiac vortical flow evolution during E-filling could allow deriving a novel method to assess MV-LV hemodynamic coupling in the presence of normal and pathologic valvular and ventricular dynamics.

In this work, we aimed to employ 4D Flow MRI to 1) Elucidate the dynamics of vortical flow within the human LV over E-filling in healthy subjects 2) Derive a standardized quantitative profile characterizing the MV-LV hemodynamic coupling during E-filling using 3D vortical flow evolution. 3) Probe the feasibility of the derived profile in detecting differences between healthy subjects and five pilot patients with various MV and LV abnormalities.

**Materials and Methods**

***Study Population***

The study was conducted across two centers, Leiden University Medical Center, the Netherlands and University of Leeds, UK, and was approved by the institutional Medical Ethical Committee in Leiden and the National Research Ethics Service in Leeds. Written informed consents were obtained from all participants or their parents.

A total of 39 subjects were retrospectively enrolled in this study; 34 healthy volunteers (age 27±16 years) and 5 example patients as representatives of valvular and/or ventricular disease: a hypertrophic cardiomyopathy (HCM) patient, a patient with restricted MV opening, a patient with remodeled LV manifested in LV dilation after myocardial infarction (MI), a patient with a presentation of both MV restriction and LV dilation, and a dilated cardiomyopathy (DCM) patient.

*Correspondence to: mohammed.elbaz@northwestern.edu

These patients were chosen to show the feasibility of the derived method in detecting differences in the presence of example MV and LV abnormalities compared to the studied healthy volunteers.

*MRI acquisition*

All subjects underwent whole-heart 4D Flow MRI on a 3T or 1.5T MRI scanner (Philips Healthcare, Best, The Netherlands) using a combination of FlexCoverage Posterior coil in the table top with a dStream Torso coil, providing up to 32 coil elements for signal reception. Velocity encoding (VENC) of 150cm/s was used in all three directions, with reconstructed spatial resolution of 2.3mm×2.3mm×(3.0mm-4.2mm), flip angle 10°, echo time (TE) 3.2ms and repetition time (TR) 7.7ms, resulting in a maximal true temporal resolution of 4×TR=31ms. Retrospective ECG-gating was used and 30 cardiac phases were reconstructed to represent one average heartbeat. To accelerate acquisition, parallel imaging SENSE with factor 2 was used and echo planar imaging with factor 5. To allow a reasonable acquisition time (8-10 minutes), free breathing was allowed for both patients as well as controls. Standard short-axis anatomical MRI cine data was as also acquired. A typical volume for a whole-heart 4D Flow MRI acquisition was 396 mm (right-left) × 336 mm (anterior-posterior) × 117 mm (feet-head). Concomitant gradient correction and local phase correction were applied using the software available on the MRI system. For more details on the MRI protocol, readers are referred to our previous publications (4, 7, 8).

*Image analysis workflow*

Using in-house developed MASS software (Leiden University Medical Center), the LV endocardial contours were segmented from 4D Flow MRI as previously described (7). In short, to tackle the typical low contrast in 4D Flow MRI that makes its LV segmentation challenging,

*Correspondence to: mohammed.elbaz@northwestern.edu

LV contours were first manually traced on anatomical short-axis slices over all acquired time phases. Contours were then projected on the whole-heart 4D Flow data. To reduce possible translation and rotation errors between cine short-axis and whole-heart 4D Flow acquisitions, automated image registration by mutual information using Elastix (18) was performed between cine short-axis data and velocity magnitude reconstructed images of the 4D Flow data using a single phase that visually showed the best depiction of the LV in the velocity magnitude 4D Flow image. Registration was restricted to translation and rotation. The resulted transformation matrix was then propagated to all 4D Flow phases. The 3D velocity field data within the segmented 4D Flow LV volume was then used to compute enstrophy parameters as described below. Segmented LV endocardial boundaries were also used to compute the LV end-diastolic volume (EDV), end-systolic volume (ESV), stroke volume (SV) and ejection fraction.

Retrospective mitral valve tracking was used to quantify inflow and derive flow-time curves from the 4D flow data as previously described (19). The time points corresponding to start of E-filling, peak early filling (E-peak) and end of E-filling were defined from the flow-time curves. For further details on our used image analysis framework and workflow, readers are referred to our previous publications (4, 7).

***Computation of vortical flow strength by enstrophy maps from 4D Flow MRI***

4D Flow MRI-derived velocity field was used to compute maps of the local strength of the vortical blood flow in the LV quantified in terms of voxel-wise enstrophy maps. Enstrophy is the squared magnitude of the vorticity vector and thus can be computed directly from velocity gradients obtained from in-vivo 4D Flow MRI. Enstrophy is an important fluid mechanics quantity as it provides a measure of the local strength of rotational motion in the flow and has been used to characterize the evolution of a variety of vortical flows and quantify energy dissipation (20).

*Correspondence to: mohammed.elbaz@northwestern.edu

4D Flow MRI provides in-vivo measurements of the three-directional velocity field $V(u, v, w)$, in the three principal directions $x, y, z$ and over the complete cardiac cycle (18). Given the acquired velocity field, the enstrophy ($\epsilon$) at a voxel (i,j,n) in the domain of interest (segmented LV volume) at time instance $t$ can be computed from the Vorticity ($\omega$) field

$$\epsilon_{i,j,n}(t) = \frac{1}{2}\omega_{i,j,n}^2(t), \omega = \left(\frac{\partial w}{\partial y} - \frac{\partial v}{\partial z}, \frac{\partial u}{\partial z} - \frac{\partial w}{\partial x}, \frac{\partial v}{\partial x} - \frac{\partial u}{\partial y}\right), i = 1:X; j = 1:Y \text{ and } n = 1:N \quad [1]$$

With $X$ as the image width, $Y$ as the image height and $N$ as the number of slices. This provides an enstrophy map at time instance $t$ over the segmented LV volume.

### Computation of instantaneous total enstrophy in the LV (TED$_{LV}$) over E-filling

To quantify the overall evolution of LV vortical flow patterns, we introduce the instantaneous non-dimensional Total (volume-integrated) Enstrophy Density (TED$_{LV}$) derived as follows:

The total enstrophy ($E_t$) at time instance $t$ over the 3D segmented LV volume of $M$ voxels and with voxel size $L$ is computed as:

$$E_t = \sum_{k=1}^{M} \epsilon_k(t) L \quad [\text{m}^3/\text{s}^2] \quad [2]$$

such that $t = 0 : T$ correspond to the acquired 4D Flow MR time frame that are spanning from the start of E-filling ($t = 0$) to the end of E-filling by the end of diastasis ($t = T$). This results in a profile curve representing the dynamic evolution of total enstrophy in the LV over E-filling. Since our control volume (the LV volume) changes dynamically over the cardiac cycle, we use the spatially-averaged enstrophy density to account to this volumetric variability (20). Accordingly, volume-normalized enstrophy density ($R$) is defined as:

*Correspondence to: mohammed.elbaz@northwestern.edu

$$R_t = \frac{E_t}{vol(LV_t)} \qquad [1/s^2] \qquad [3]$$

such that $vol(LV_t)$ is the volume of the segmented LV region of interest at acquired time frame $t$. Of note, $R_t$ computation involves normalization by instantaneous LV volumes. Therefore, it accounts for the potential difference in LV size between different subjects and also takes into consideration the LV size change over the diastolic filling of the same subject.

To further enable a standardized analysis over different subjects, it is best to evaluate vortical flow profile in a standardized dimensionless means. Therefore, to account for subject-specific variability of the vorticity field, the so obtained enstrophy integral ($R_t$) is normalized by the time-peak enstrophy for this individual over the entire E-filling process (i.e. instantaneous TED$_{LV}$ values are time-normalized relative to time-peak TED$_{LV}$ — time-peak TED$_{LV}$ has normalized to 1):

$$TED_{LV}(t) = \frac{R_t}{\max_{1 \leq r \leq T} R_r}, \; t = 0:T, \; \text{[dimensionless]} \qquad [4]$$

$T$ is number of acquired E-filling phases as described above. Using this normalization approach allows standardized comparison of $TED_{LV}$ between different patients.

### Construction of standardized MV-LV vortical flow evolution profile (P$_{MV-LV}$)

To quantify vortical flow dynamics in the LV over E-filling and to link to both MV and LV hemodynamics, we constructed a temporal vortical flow profile (P$_{MV-LV}$) as a function of both MV and LV vorticity dynamics over E-filling. The dimensionless profile P$_{MV-LV}$ is constructed from Equation 4 and 5 by plotting LV enstrophy density evolution (TED$_{LV}$) as the y-axis relative to the temporal vorticity discharge rate from MV by means of the dimensionless temporal vortex formation time ($T_{MV}$) as the X-axis. Following (15), $T_{MV}$ is computed as:

$$T_{MV_t} = \int_0^z \frac{U}{D} \, dt, z = t * \Delta t, t = 0: T \quad \text{[dimensionless]} \quad [5]$$

*Correspondence to: mohammed.elbaz@northwestern.edu

$\Delta t$ is the temporal resolution of acquired 4D Flow MRI, $U$ is the instantaneous inflow velocity through the MV and $D$ as maximum diameter of mitral inflow opening. D was computed at E-peak from the area of the MV flow on retrospective valve tracking (19) at peak inflow velocity level (i.e., approximately at the tip of the valves), assuming a circular inflow area. In the original definition, D=constant. However in our case, we use D = D(t) since the valve opens and closes over E-filling. This is similar to the concept used by Gharib et al in (1) with D defined as the maximum MV inflow diameter over E-wave. We used the maximum diameter ($D_{max}$) instead of average diameter as in (1) because we estimate MV dimeter from flow data in which MV region can be unclear near to diastasis when little or no inflow is present. As such, using $D_{max}$ assures more reliable measurement of MV diameter derived from 4D Flow MRI.

As such, $P_{MV-LV}$ characterizes the MV and LV hemodynamic coupling through vortical flow dynamics by linking the enstrophy density derived from LV vorticity dynamics to the vortex formation time derived from MV vorticity dynamics. Note that $P_{MV-LV}$ is a dimensionless profile. Thus, facilitates standardized comparison of MV-LV hemodynamic coupling between different patients.

All enstrophy analyses were performed using an in-house software module developed using MATLAB (MathWorks Inc.).

8
*Correspondence to: mohammed.elbaz@northwestern.edu

*Statistical Analysis*

Data analysis was performed using SPSS Statistics (version 20.0 IBM SPSS, Chicago, IL). Variables were tested for normal distribution using the Shapiro-Wilk test. P<0.05 was considered statistically significant. Continuous variables are expressed as mean ± standard deviation (SD) or as median with inter-quartile range (IQR) where appropriate.

Table 1. Characteristics of healthy volunteers and patients.

|  | Healthy volunteers (N=34) | DCM patient | HCM patient | MI patient | Restricted Mitral valve patient with normal size LV | Restricted Mitral valve patient with dilated LV size |
|---|---|---|---|---|---|---|
| **Age, years** | 27±16 | 73 | 65 | 71 | 12 | 16 |
| **Sex (Female(F)/Male(M))** | 16 F | M | F | M | F | F |
| **Heart rate, bpm** | 68±17 | 64 | 90 | 80 | 114 | 60 |
| **LVEDV (ml)** | 147±40 | 279 | 147 | 200 | 129 | 215 |
| **LVESV (ml)** | 57±19 | 226 | 79 | 138 | 62 | 92 |
| **SV (ml)** | 90±23 | 53 | 67 | 61.5 | 67 | 123 |
| **BMI (kg/m$^2$)** | 21±4 | 26 | 24 | 33.8 | 44 | 29 |
| **LV Mass (g)** | 83±24 | 130 | 121 | 104 | 68 | 91 |
| **Ejection Fraction, %** | 62±4 | 19 | 46 | 30.8 | 52 | 57 |
| **MV peak inflow area (cm$^2$)** | 9.66±2.47 | 14.91 | 5.66 | 17.95 | 5.58 | 5.17 |

LV= Left Ventricle, LVEDV= Left Ventricular End Diastolic Volume, LVESV= Left Ventricular End Systolic Volume, SV=Stroke Volume, BMI=Body Mass Index, DCM= LV Dilated Cardiomyopathy, HCM=LV Hypertrophic Cardiomyopathy. MI=Myocardial Infarction, ± refers to standard deviation

*Correspondence to: mohammed.elbaz@northwestern.edu

**Results**

*Clinical Characteristics*

Clinical characteristics of the studied subjects are summarized in Table 1.

*Bi-phasic vortical flow process regulated by MV-LV hemodynamic coupling*

Figure 1, and Supplementary Video 1, show an example of the 4D Flow MRI-derived instantaneous enstrophy maps, computed over all E-filling frames (12-frames) of the acquired 30 time-frame cine 4D Flow MRI in a 29-year-old male visualized in a 4 chamber and (mid-) basal short axis cross-sectional views. The overall evolution of large-scale vortical flow patterns in the LV over diastole as revealed by the enstrophy maps in Figure 1 was found to be consistent among all the studied healthy subjects. With the start of MV opening, vortical flow (marked by the regions of high enstrophy density) emerges distally to the mitral valve near the leaflets tip. Consistent with previous reports, this vortical flow continues to grow while organizing into a ring-like structure near to the moment of the peak E-filling (1, 3-5, 14). However, following peak E-filling, this vortex ring propagates at an inclined angle towards the LV free wall giving rise to vortex-wall interaction. At a time point past-peak E-filling, the vortex ring flow starts to decay becoming disorganized as it gradually loses its distinct ring-like shape. We note that this apparent decay phase of the vortex ring emanating from the MV has not been reported before in-vivo though it has been postulated by numerical simulation (3).

*Correspondence to: mohammed.elbaz@northwestern.edu

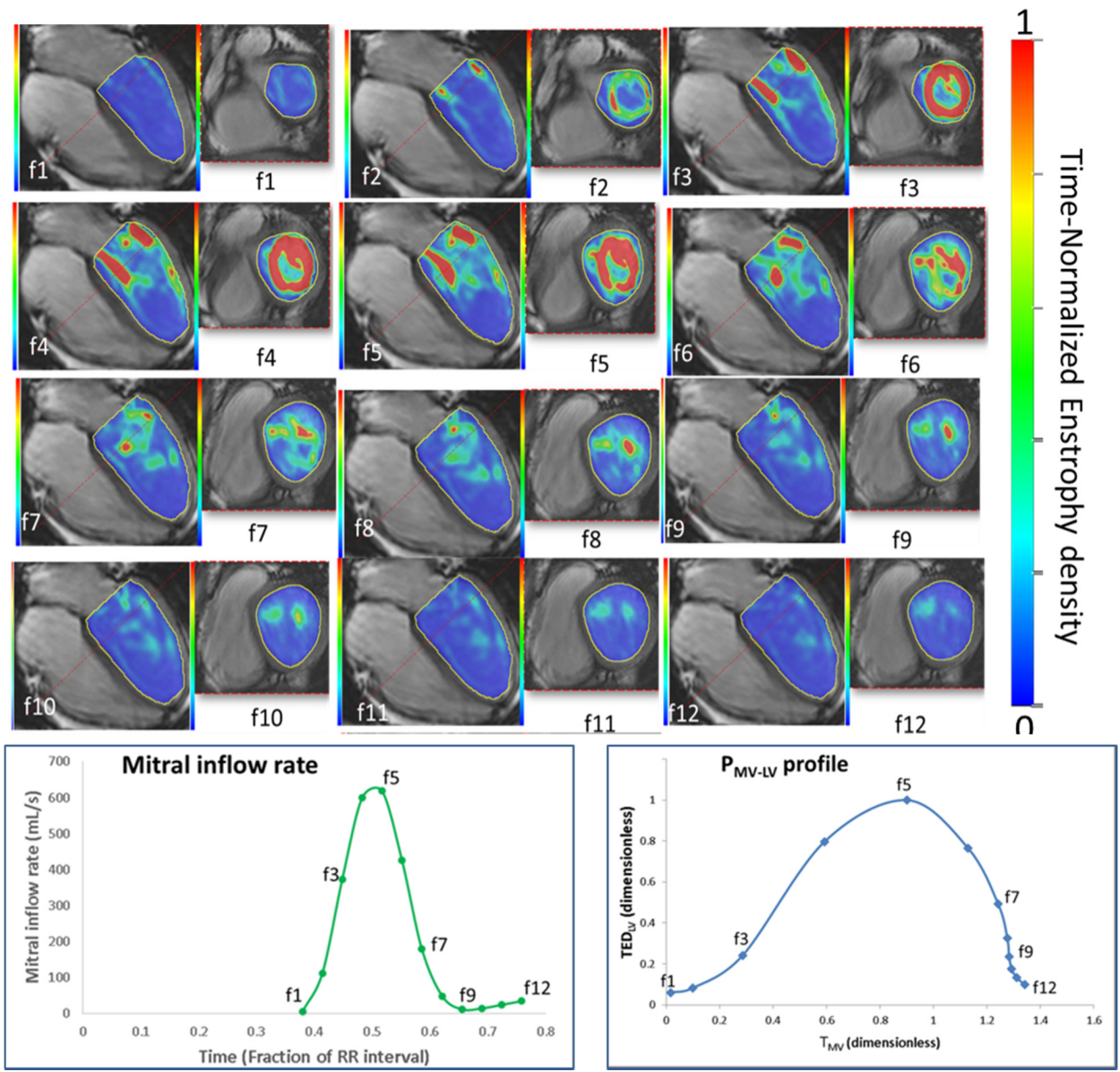

**Figure 1: Vortical flow evolution in the LV over E-wave by means of enstrophy maps in a healthy 29-year-old volunteer**: Enstrophy maps are color coded from blue (minimum) to red (maximum). Presented enstrophy values are dimensionless (normalized to temporal peak enstrophy value). For each acquired E-wave time-phase, from the computed 3D enstrophy maps, two standard cross-sectional views are presented: a four-chamber view (on the left) and next to it is a short-axis view (on the right) at the mid-basal level highlighted by the dotted red line. The curve on bottom left represents the MV inflow rate from which the start and end of filling phases are derived. The dots represent the acquired 4D Flow MRI time-phases. In this subject, E-filling is defined from start of E-filling at time-point (f1) until the end of diastasis at time-point (f12). The profile on the bottom right shows the derived dimensionless $P_{MV-LV}$ profile corresponding to this subject.

*Correspondence to: mohammed.elbaz@northwestern.edu

The enstrophy in the flow continues to decay during E-inflow deceleration until it essentially diminishes by the end of diastasis (Figure 1). No pockets of intense vorticity (high enstrophy levels) are seen to be able to reach the apex by the end of the E-wave. These results show that MV inflow dynamics gives rise to the formation of the vortical flow (including vortex ring) and impacts its growth (1, 3, 4) while the LV wall dynamics governs its subsequent decay. Therefore, this supports our hypothesis that vortical flow evolution in the LV is not only a function of MV inflow but rather regulated by a hemodynamic coupling between MV inflow and LV wall geometry dynamics.

**Standardized dynamic vortical flow profile ($P_{MV\text{-}LV}$) *and its characteristic parameters***

Figure 2a shows a representative example of the dynamic vortical profile $P_{MV\text{-}LV}$ of a healthy volunteer and presents definitions of the derived characteristic parameters. Table 2 presents the results of the derived characteristic parameters over all studied healthy subjects. In line with our reported vortical flow evolution by enstrophy mapping above, the derived $P_{MV\text{-}LV}$ profile of all healthy subjects (Figure 2a, 2b) exhibits two distinct phases: the vortical flow growth phase and the vortical flow decay phase. The vortical flow growth phase starts with the opening of MV generating enstrophy along the shear layers emanating from the MV leaflets (Figure 1). The $TED_{LV}$ reaches its peak at a dimensionless vortical formation time $T_{growth} = 1.23 \pm 0.25$ with a growth rate (slope) $\alpha = 0.80 \pm 0.17$, which marks the end of the vortical growth phase. Following the peak, the $TED_{LV}$ starts to decay, as vortex structures begin to interact with the LV wall (see Figure 1), with total decay time $T_{decay} = 0.96 \pm 0.39$ and decay rate (slope) $\beta = -1.02 \pm 0.49$. Note that at the start of inflow discharge, the value of $TED_{LV}$ was $TED_{start} = 14 \pm 9\%$ of the time-peak $TED_{LV}$ value while at the end of E-filling $TED_{end} = 18 \pm 9\%$ of the time-peak $TED_{LV}$ value.

*Correspondence to: mohammed.elbaz@northwestern.edu

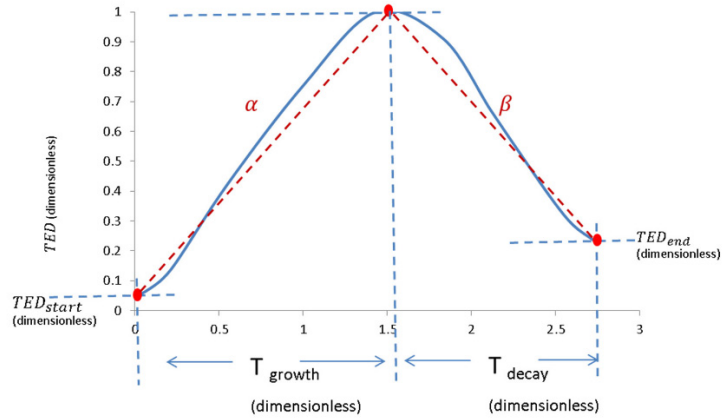

(a)

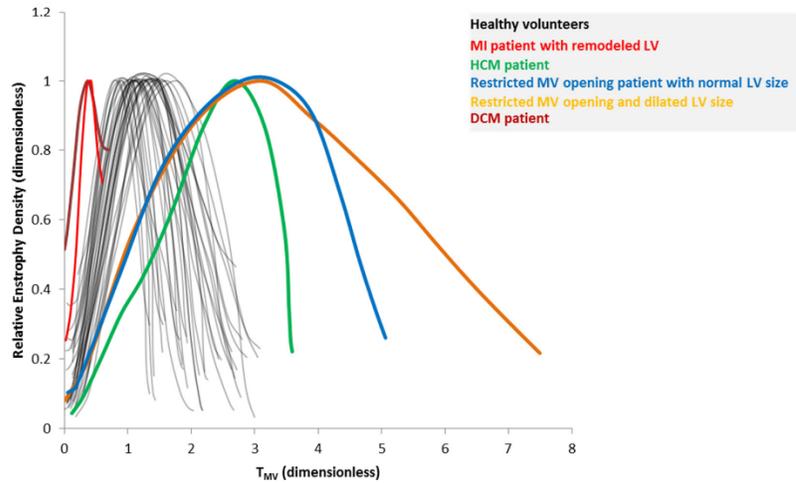

(b)

**Figure 2: Parameterized dimensionless temporal vortical blood flow profile ($P_{MV\text{-}LV}$)** (a) An example temporal $P_{MV\text{-}LV}$ profile of a healthy volunteer demonstrates the different characteristic parameters of the evolution of total dimensionless enstrophy in the LV over the dimensionless vortex formation time($T_{MV}$). From the enstrophy evolution curve two main phases are defined, the vortical formation phase can be characterized by the following parameters: the non-dimensionless vortical growth duration $T_{growth}$, the enstrophy growth rate α and $E_{start}$ which is the enstrophy level in the LV at the start of MV discharge as percentage of the time-peak enstrophy level. The vortical decay phase can be characterized by the three parameters: the non-dimensionless duration of vortical flow decay $T_{decay}$, the enstrophy decay rate β and $E_{end}$ which is enstrophy level in the LV at the end of E-filling as percentage of the time-peak enstrophy level. (b) shows the significantly altered $P_{MV\text{-}LV}$ profiles in the 5 sample cardiac patients studied (in color) compared to $P_{MV\text{-}LV}$ profiles of all 34 healthy volunteers studied (in grey). Note that the $P_{MV\text{-}LV}$ profiles of all healthy volunteers present the same behavior and collapse within a narrow band of variability.

*Correspondence to: mohammed.elbaz@northwestern.edu

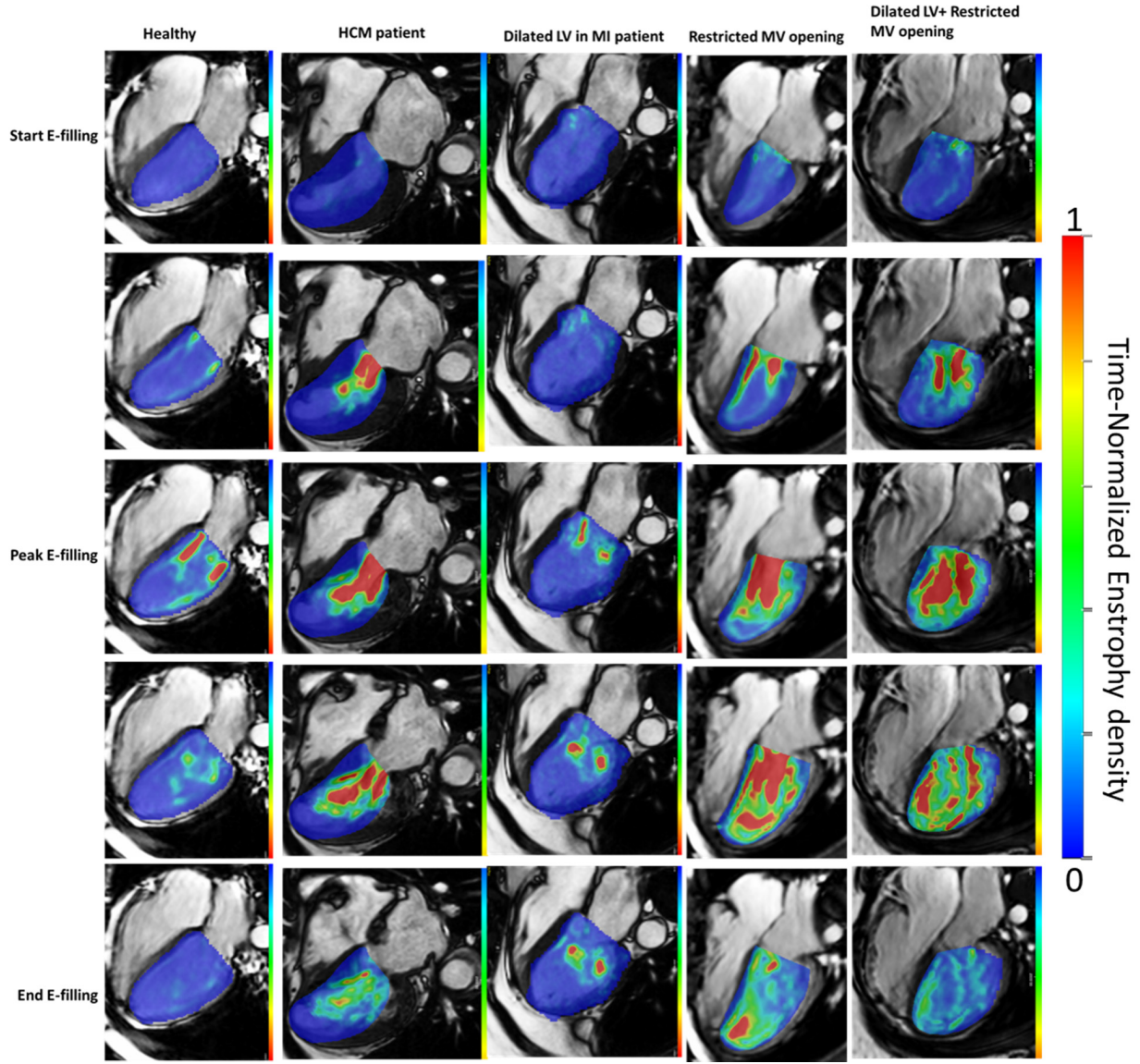

**Figure 3: Temporal enstrophy maps in the LV over E-filling in sample 4 cardiac patients.** For visual clarity, only 2D cross-sectional views in a standard four-chambers view are presented out of the computed 3D time-varying enstrophy volume maps. Enstrophy maps are presented over 5 sample time-points spanning over the E-filling duration. Please note that the quantitative results of vortical flow profile are computed over the entire 3D time-varying enstrophy maps. As in Figure 1, presented enstrophy values are dimensionless. All enstrophy scales are normalized to the scale of the healthy volunteer to provide the same color scale for all presented cases. Color coding from blue (minimum) to red (maximum).

14
*Correspondence to: mohammed.elbaz@northwestern.edu*Correspondence to: mohammed.elbaz@northwestern.edu

### Distinct $P_{MV-LV}$ profile signatures of MV versus LV abnormalities in the studied example patients

Figure 2b shows the $P_{MV-LV}$ profiles corresponding to the 5 representative patients as compared to $P_{MV-LV}$ profiles of all studied 34 healthy volunteers. Figure 3 presents the computed enstrophy maps of these patients in a standard four chambers cross-sectional view of the heart at five sample time points over E-filling (Supplementary videos 3-7 show evolution over complete E-filling). Table 2 presents a summary of the vortical profile's characteristic parameters for each of the individual patients studied. As hypothesized, results reveal distinct and readily quantifiable differences in the $P_{MV-LV}$ vortical profiles between patients and healthy volunteers.

**Table 2.** Characteristic dimensionless parameters of the dynamic vortex profile ($P_{MV-LV}$) in healthy volunteers (N=34) and individual representative patients

|  | $T_{growth}$ | growth rate ($\alpha$) | $T_{decay}$ | decay rate ($\beta$) | $TED_{start}$ | $TED_{end}$ |
|---|---|---|---|---|---|---|
| **Healthy volunteers (N=34)** | 1.23±0.25 | 0.80±0.17 | 0.96±0.39 | -1.02±0.49 | 14 ± 9% | 18 ± 9% |
| **HCM Patient** | 2.69 | 0.37 | 0.90 | 0.86 | 4% | 22% |
| **MI patient** | 0.42 | 1.88 | 0.18 | -1.65 | 25% | 71% |
| **Restricted Mitral valve patient with dilated LV size patient** | 3.08 | 0.30 | 4.41 | -0.18 | 8% | 22% |
| **Restricted Mitral valve patient with normal size LV patient** | 2.79 | 0.33 | 2.27 | -0.33 | 10% | 26% |
| **DCM patient** | 0.37 | 1.33 | 0.34 | -0.59 | 52% | 80% |

*Correspondence to: mohammed.elbaz@northwestern.edu

**Discussion**

Using in-vivo 4D Flow MRI-derived dynamic enstrophy mapping, this study elucidates a bi-phasic vortical flow evolution over the E-filling in the healthy LV. This bi-phasic behavior is characterized by a vortical growth phase followed by a vortical decay phase. To our knowledge this is the first in vivo study to use enstrophy to characterize vortical flow evolution in the LV. A standardized profile of vortical flow evolution is derived as a function of LV enstrophy density relative to the MV vorticity discharge over the E-wave. The derived profile parameters quantitatively characterize the revealed bi-phasic behavior of vortical flow evolution by means of coupled MV-LV hemodynamics. Hence, enabling a comprehensive analysis of diastolic vortical flow evolution. Consistent vortical evolution ($P_{MV-LV}$) profile parameters are found over all healthy volunteers and distinct alterations observed in the example patients studied.

***Bi-phasic vortical flow profile and relation to MV-LV hemodynamic coupling in healthy volunteers***

Previous studies showed that vortical flow emerges from the MV (1, 2, 14, 21, 22) and progresses inside the LV over E-filling (3, 4, 6, 11, 16, 17, 23). As such, vortical flow dynamics over E-filling could serve as a linkage between MV and LV hemodynamics. Therefore, in this work, we aimed to derive a dimensionless profile that links both MV and LV hemodynamics through vortical flow evolution to allow a standardized analysis of MV-LV hemodynamic coupling. To achieve such profile, we needed to define a standardized measure of vortical flow dynamics from the MV and link it to an appropriate measure of vortical flow evolution from the LV. Importantly, such measures needed to be time-varying and able to characterize fluid-dynamic properties of vortical flow including formation and dissipation. With regard to MV vortical dynamics, we used vortex formation/progression time (indicated here as $T_{MV}$), previously introduced by Gharib et al (1, 15) that characterizes the vorticity discharge from MV as function of the MV geometry (diameter) and the dynamic inflow velocity over E-filling. Hence, providing an appropriate time-varying index of vorticity discharge from MV over E-filling that also serves as a dimensionless time-scale (15). With respect to LV vortical flow dynamics, we used enstrophy mapping of the 4D Flow MRI-derived flow field. Enstrophy is an important fluid dynamics quantity that is known to characterize vortical flow evolution and dissipation (20, 24, 25). Therefore, enables characterization of the complex dynamic vortical flow behavior in the LV. For an intuitive analogy, relation of enstrophy to vorticity (curl of velocity) is analogous with that of kinetic energy to velocity i.e. enstrophy could be, loosely, thought of as a measure of vortical flow energy therefore able to characterize growth and dissipation/decay of vortical flow.

*Correspondence to: mohammed.elbaz@northwestern.edu

In the healthy LV, prior to peak E-filling, results of our enstrophy maps are consistent with previous studies of vortical flow formation in the LV including the formation of a vortical ring structure (1, 3-5, 14). However, following peak E-filling, a major finding of the enstrophy maps in this study is the observation of a vortical decay phase in all healthy volunteers studied that has not been characterized before in vivo. Therefore, our results in the healthy volunteers reveal that that the evolution of vortical flow over E-filling is not only dominated by the formation of the vortical flow from the MV (1, 21), but rather has a more complex bi-phasic behavior involving growth and decay of the vortical flow. These findings support our hypothesis that vortical flow evolution in the LV is a result of a hemodynamic coupling between MV inflow and LV wall geometry dynamics with MV inflow dynamics gives rise to the formation of the vortical flow and impacts its growth (1, 3, 4) while LV wall dynamics governs its following decay.

Notably, our in-vivo results support the recent computational and in-vitro experimental studies that indicated that the ensuing interaction of vortical flow with the surrounding dynamically evolving LV geometry during diastole influences the temporal evolution of the vortical blood flow (3, 11, 13). Our results show that, with $T_{growth}$ =1.23±0.25 vortical flow in the healthy LV stops its growth phase at a time-point near peak E-filling (start of inflow deceleration).

*Correspondence to: mohammed.elbaz@northwestern.edu

Recent in-vivo studies (16, 17) showed, using Lagrangian Coherent Structures technique (LCS), that the LV vortex boundary follows the endocardial boundary which agrees with our results herein. While these studies provided a better understanding of LV vortical formation, it is not straightforward to compare LCS dynamics across subjects. We extend the knowledge from previous studies by providing a standardized method to quantify the evolution (not only formation) of vortical flow and linking it to MV-LV hemodynamics among studied subjects. We showed herein that the previously reported proximity of vortical flow to the LV boundary in the healthy LV (16, 17) induces a vortical decay phase that has not been reported before. A potential explanation of the discrepancy on the revelation of the decay phase in our study as compared to previous studies could be, in part, due to the methodology used and the focus of the study. In the majority of previous studies the focus had been on characterizing the vortical formation properties and not on the entire evolution as in this work. Consequently, previous studies have used methods that are more adequate to identify or characterize vortical flow formation (e.g. LCS or optimal vortex formation time) (1, 16, 17) but not necessarily adequate or meant for characterizing vortical flow evolution in the entire diastolic E-filling. In this study, we used enstrophy mapping which is known to be adequate to characterize vortical flow evolution properties including formation and dissipation (20, 26) to serve our focus on studying the evolution of vortical flow in the LV.

### Vortical $P_{MV-LV}$ profile over E-filling: distinct signatures beyond vortex formation characteristics

While previous in-vivo vortex quantification parameters have been mainly devoted to characterization of vortex formation properties (1, 5, 6), our derived $P_{MV-LV}$ profile allows quantitative characterization of different aspects of vortical flow evolution over the entire LV. It is important to highlight that unlike vortex identification-based analysis (4, 16, 17) that requires data thresholding to identify structures of interest, our enstrophy profile is threshold-free as it characterizes the entire vortical flow in the LV. Being dimensionless and threshold-free, the derived $P_{MV-LV}$ profile allows for a standardized comparison between vortical flow dynamics and link to MV-LV hemodynamics between different subjects in health and disease. This is a critical property of $P_{MV-LV}$ profile, which could allow for assembling the hemodynamic characteristics in a large number of patients in a standardized quantitative manner. Therefore, $P_{MV-LV}$ profile may potentially allow in the future to identify early blood flow markers in heart disease.

*Correspondence to: mohammed.elbaz@northwestern.edu

Our pilot results in the studied patients clearly show the feasibility of the derived vortical profile in detecting distinct quantifiable differences in $P_{MV-LV}$ profile parameters depending on the cardiac abnormality as compared to all healthy volunteers studied. Importantly, results noticeably show that cardiac valvular and/or ventricular abnormalities do not only affect the formation/growth phase of vortical flow as has been the main focus of previous research (1, 4-6). Instead, valvular and or ventricular abnormalities are shown to affect different characteristics of vortical flow evolution manifested in altered $P_{MV-LV}$ growth and decay parameters. Enstrophy maps show a considerably increased enstrophy density in the mitral inflow discharge in the HCM patient, restricted MV patient with dilated LV size and restricted MV patient with relatively normal LV size (Figure 3 Supplementary Movie 2, 4, 5). That is quantitatively reflected in the $P_{MV-LV}$ profile of these patients by a considerably prolonged (2-3 time) $T_{growth}$ compared to healthy volunteers. Notably, while these patients show similar growth parameters, their decay parameters, which cannot be inferred from formation properties alone, differ substantially leaving each of the patients with a distinct $P_{MV-LV}$ profile signature.

Both MI and DCM patients show a reduced vortical growth time. Nevertheless, these patients show distinctly altered decay properties from healthy volunteers and from each other (Figure 3, Supplementary Movie 3, 6). A $TED_{end}$ of 71% in the MI patient and 80% in the DCM patient indicate that the vast majority of the discharged vortical flow is retained in the ventricle by the end of E-filling demonstrating complex vortical flow in the ventricle. A $TED_{start}$ of 25% in the MI patient and 52% in the DCM patient indicate that large amount of vortical flow density has retained in the ventricle at end-systole which might contribute to increased levels of hemodynamic energy losses (7, 9).

*Correspondence to: mohammed.elbaz@northwestern.edu

While our feasibility results are clearly encouraging, caution need to be taken in interpreting such early results or extrapolating them to other patients. Follow-up cohort-based studies are needed to further confirm such results and to understand the link to cardiac function.

### *Potential Clinical implications*

Altered LV or MV geometrical or dynamic properties due to heart disease would impact the vortical flow evolution as reflected in altered characteristics of the derived $P_{MV-LV}$ profile relative to that of the healthy heart. Mitral valvular impairments or deformations would adversely impact the characteristic of vortical growth parameters in the $P_{MV-LV}$ profile in a similar manner to that exemplified in the pilot patients studied herein i.e. the HCM patient and the two patients with restricted MV opening (Figure 3). Meanwhile, LV geometry deformations including myocardial stiffness, impaired relaxation or wall motion abnormalities such as those seen in cardiomyopathy(27), physiological LV remodeling as in healthy athletes or pathological remodeling such as in patients after myocardial infarction or LV hypertrophy(28) would alter the flow-structure interaction potentially affecting vortical decay behavior as reflected in altered $P_{MV-LV}$ profile decay parameters in a similar manner to that shown in the example MI and DCM patients explained above. A combination of valvular and ventricular abnormalities can be present in different heart diseases as in LV remodeling after ischemic mitral regurgitation (29), cardiomyopathy patients (30) or diastolic dysfunction (31). Such combined MV-LV abnormalities could be quantitatively characterized by $P_{MV-LV}$ profile as exemplified in the example patient herein presented with both restricted MV and LV dilation showing distinct alteration in both the growth and decay parameters of $P_{MV-LV}$ profile. As such, intra-cardiac vortical flow evolution as characterized by the $P_{MV-LV}$ profile could enable a deeper understanding of the association between valvular and ventricular health in different heart diseases. Our results could pave the way to better detection tools of the coupling between valvular and ventricular remodeling and progression as reflected in vortical flow evolution. Furthermore, provided vortical flow analysis methodology could lead to an objective assessment of the effectiveness of treatment strategies for cardiac diseases (e.g. Mitral valve repair)(32, 33). Future studies are needed, however, to reveal how specific pathophysiological events in different cardiac diseases affect vortical flow evolution and association to cardiac function. Moreover, our revealed bi-phasic vortical flow dynamics can be an important step to guide the design of accurate in-vitro experiments and numerical in-silico patient-specific simulations that resemble realistic cardiac hemodynamics(33).

*Correspondence to: mohammed.elbaz@northwestern.edu

## *Limitations*

A limitation of this study is relatively small number of subjects. However, our results show consistent behavior with a narrow band of variability in all studied healthy subjects which could further emphasize the reliability of our results. While only five example patients were studied, the aim was not to provide health outcomes of these patients. Instead, these patients were carefully chosen with various MV and LV abnormalities to experimentally illustrate the feasibility of the method to pave the way for future cohort-specific studies using our methodology for investigating MV-LV hemodynamic coupling and its relation to cardiac function. The derived $P_{MV-LV}$ profile can be considered as a global quantification method i.e. it shows that different enstrophy density is present but does not identify the spatial distribution of enstrophy within the LV. However, for a localized analysis, the global-based analysis by $P_{MV-LV}$ profile could be readily accompanied by the presented enstrophy maps to identify enstrophy density in different regions of the heart. 4D Flow MRI was acquired without respiratory gating to allow a reasonable acquisition time (8-10 minutes). We have used this protocol in several of our previously published work (4, 7, 8). It should be acknowledged, however, that lack of respiratory gating could potentially result in blurring in the acquired flow field and related errors. We have visually inspected the data and we have not identified any significant blurring. Furthermore, a recent publication showed that 4D flow MRI acquired without respiratory gating yields comparable quantitative measurements of vortex formation to 4D flow MRI with respiratory gating (34).

*Correspondence to: mohammed.elbaz@northwestern.edu

4D Flow MRI includes averaging the flow field data over several cardiac cycles. This time-averaging would potentially smooth out the small-scale flow features and does not, generally, adjust for flow deviations resulting from beat to beat variability. However, it was the aim of this study to evaluate large-scale cardiac vortical flow patterns defined as being with larger size than the acquired voxels and with a longer time-span than the acquired temporal resolution from 4D Flow MRI (in this study a spatial resolution of 2.3mm×2.3mm×3mm-4.2 mm and a temporal resolution of 31ms were used). Vortical flow features beyond acquired resolution might not be negligible but could not be captured and therefore not included in the provided analysis.

In conclusion, using 4D Flow MRI, we derived a standardized profile that quantitatively characterizes MV-LV hemodynamic coupling through vortical flow dynamics by means of LV enstrophy mapping and MV vorticity discharge. This profile shows consistent bi-phasic vortical evolution behavior in healthy subjects characterized by a growth phase followed by a decay phase. Distinct quantifiable alterations were found in the derived profile of the pilot example patients studied with MV and LV abnormalities. These findings unravel new insights into human heart's physiology and could enable a novel method to understand the coupling between the valvular and ventricular hemodynamics and its association with different heart diseases. Future studies with large cohorts are needed to reveal how specific pathophysiological events in different cardiac diseases affect vortical flow evolution and association to cardiac function.

*Correspondence to: mohammed.elbaz@northwestern.edu

*Correspondence to: mohammed.elbaz@northwestern.edu

*Correspondence to: mohammed.elbaz@northwestern.edu